\documentclass[runningheads]{llncs}
\usepackage[T1]{fontenc}
\usepackage{graphicx}
\begin{document}
\title{Two Explorative Studies on Tangible Augmented Reality for
Neurodevelopmental Disorders}
\titlerunning{Two Explorative Studies on TAR for NDDs}
%
\author{Francesco Vona\inst{1}\orcidID{0000-0003-4558-4989} \and
Giulia Valcamonica\inst{2}\orcidID{0009-0007-0089-1594} \and
Franca Garzotto\inst{2}\orcidID{0000-0003-4905-7166}}
\authorrunning{F. Vona et al.}
%
\institute{University of Applied Sciences Hamm-Lippstadt, Germany
\email{name.lastname@hshl.de}\\ \and
Politecnico di Milano, Italy \email{name.lastname@polimi.it}}
\maketitle              
\begin{abstract}
Tangible Augmented Reality (TAR) is an interaction paradigm that integrates physical and digital worlds to create immersive, interactive experiences. This paper explores two TAR applications—Holomarket and Along the Oceanic Flow (ATOF)—and presents insights from two exploratory studies evaluating their usability and likeability among individuals with NDD.
Holomarket is designed to simulate a supermarket shopping experience, helping users develop essential life skills such as item selection, basic arithmetic, and money handling. Participants interacted with augmented food items and a smart cash register, navigating a virtual supermarket environment. While participants enjoyed the realistic setting and tangible interactions, some usability challenges, such as difficulty manipulating virtual objects and discomfort with prolonged headset use, were noted.
ATOF transforms the user’s environment into an oceanic world, where participants use a dolphin-shaped Smart Object to complete tasks like collecting items and solving puzzles. This application aims to improve motor coordination and cognitive skills. Participants appreciated the immersive experience, the customizable tasks, and the tangible dolphin interface. However, some faced difficulties interacting with specific virtual elements.
Overall, both applications demonstrated potential as therapeutic tools for NDD, offering engaging, immersive experiences. Despite some usability challenges and hardware limitations, the positive feedback suggests TAR could play a crucial role in future therapeutic interventions. However, further research is needed to refine these applications and enhance user interaction and comfort.

\end{abstract}
\section{Introduction}
The rapid advancement of immersive technologies has opened new frontiers in therapeutic interventions for individuals with Neurodevelopmental Disorders (NDD). These disorders, which include Autism Spectrum Disorder (ASD), Attention-Deficit/Hyperactivity Disorder (ADHD), and Intellectual Disabilities and others, often impair cognitive, social, and motor functions, affecting a person’s ability to navigate daily life. While there is no definitive cure, therapeutic approaches often focus on rehabilitation and fostering greater independence through tailored interventions \cite{dsm}.
Among these technologies, Augmented Reality (AR) has emerged as a powerful tool by blending digital and physical environments in real time, offering interactive, three-dimensional experiences. However, AR-based therapies often lack tangible interactions, which are crucial for individuals with NDD who benefit from hands-on, multisensory engagement. To address this limitation, we exploit Tangible Augmented Reality (TAR), an innovative interaction paradigm that integrates AR with tangible interfaces, specifically Smart Objects, to create intuitive and engaging therapy experiences.
TAR enhances AR-based interactions by incorporating physical, graspable elements that provide tactile and sensory feedback. These Smart Objects, equipped with sensors and actuators, bridge the gap between digital and physical interaction, making therapy sessions more immersive, engaging, and potentially more effective. By leveraging both AR and tangible interfaces, TAR offers a novel way to enhance cognitive, motor, and social skills in individuals with NDD through meaningful, hands-on experiences.
This paper presents two TAR-based applications designed for individuals with NDD:
\begin{enumerate}
    \item Holomarket – A grocery shopping simulation developed in collaboration with psychologists and caregivers to improve memory, attention, and numerical skills. The system features holographic food shelves, a smart shopping list, and an RFID-enabled cash register, creating a structured and interactive learning environment.
    \item Along the Oceanic Flow (ATOF) – An interactive ocean-themed therapy environment, where participants engage with digital marine life and a dolphin-shaped Smart Object to improve motor coordination, cognitive abilities, and problem-solving skills through story-driven tasks.
\end{enumerate}
Both applications were evaluated in exploratory empirical studies involving individuals with NDD, focusing on usability and likeability. The results suggest that TAR has the potential to enhance therapeutic interventions by combining immersive digital content with interactive tangible elements, offering a more engaging and effective approach to therapy.

\section{Related Work}
By integrating tangible objects with Augmented Reality, TAR enables users to physically interact with digital content, making it particularly valuable in education, therapy, and design \cite{2,3,1}. This approach enhances engagement and learning by providing hands-on experiences that combine the intuitive benefits of physical interaction with the flexibility of digital augmentation. In therapeutic settings, immersive technologies like AR and Virtual Reality (VR) have already demonstrated significant potential for individuals with Neurodevelopmental Disorders, contributing to improvements in motor skills, cognitive functions, and social interactions \cite{4,5,9,7}. However, while AR and VR have shown great promise, they often lack tangible, graspable elements, which are essential for individuals who benefit from physical engagement and sensory feedback. Despite the recognized advantages of integrating Smart Objects into AR-based interventions, this aspect remains underexplored in the NDD field. The use of tangible interfaces in therapeutic applications has shown encouraging results. For example, Polipo, a tangible toy designed for children with NDD, demonstrated how physical interaction can enhance cognitive and social development through engaging, hands-on interactions \cite{12}. Additionally, systematic reviews on AR-based interventions for Autism Spectrum Disorder (ASD) have highlighted improvements in social, cognitive, and behavioral skills, reinforcing AR’s role as a valuable therapeutic tool \cite{13}. Similarly, research exploring VR and AR for individuals with communication disabilities and NDDs has emphasized their effectiveness in enhancing communication and engagement across different age groups \cite{14}. Beyond direct therapy, researchers have explored ways to enhance familiarity and comfort with AR technologies for individuals with NDD. For example, the authors of \cite{10} conducted a pilot study using the HoloLens 2 to introduce individuals with severe ASD to AR technology, leveraging eye-tracking to assess engagement and learning indicators. Likewise, the use of tangible digital storytelling tools for adolescents with NDD was also investigated by \cite{11}, demonstrating significant improvements in cognitive and social skills through collaborative storytelling. Recent technological advancements have further expanded the potential of Tangible Augmented Reality. One of these, is the concept of tangible holograms, blending virtual objects with physical augmentations to create mobile, interactive experiences that offer tactile feedback and engagement \cite{8}. Similarly, research into continuously scanning a user’s surroundings to match virtual objects with their physical counterparts has demonstrated how such solutions can reduce visual-haptic mismatches and create more immersive, realistic interactions \cite{24,6}. Additionally, capacitive 3D-printed objects have been proposed as a way to enable remote interaction while maintaining a tangible, co-located experience \cite{41}. Despite these promising advancements, few approaches have been specifically designed to address the unique needs of individuals with NDD. The integration of Smart Objects with AR environments offers an unexplored opportunity to enhance therapeutic interventions by combining immersive digital content with hands-on, tangible interactions. This study aims to fill this gap by investigating how TAR-based solutions can support autonomy, cognitive development, and social skills in individuals with NDD. By leveraging the strengths of both physical and digital interaction, TAR has the potential to redefine therapeutic experiences, offering engaging, multisensory interventions that can lead to meaningful improvements in everyday life skills.

\section{First Study - Along the Oceanic Flow}
Along the Oceanic Flow (ATOF) is an application designed to help people with NDD improve motor and cognitive skills through game exercises. Utilizing the TAR interaction paradigm \cite{Carmigniani2011-hv}, the user must complete some experiences by performing motor and cognitive tasks such as collecting elements, avoiding obstacles, and discovering a way to open treasure chests in his or her own room, which has been transformed into an ocean floor full of interactive decorations. Each task involves the user's use of a dolphin-shaped puppet, "smartified" so that it can interact with the Hololens. The tasks have been designed to be extremely simple to do, and one of them includes an experience in which the exercises gradually get harder as the game progresses. Along the Oceanic Flow was tested in an exploratory study to assess the likeability and usability of the application. The study involved 11 users with NDD and lasted two days.
\subsection{Application Overview}
\begin{figure}[ht]
    \centering
    \includegraphics[width=\linewidth]{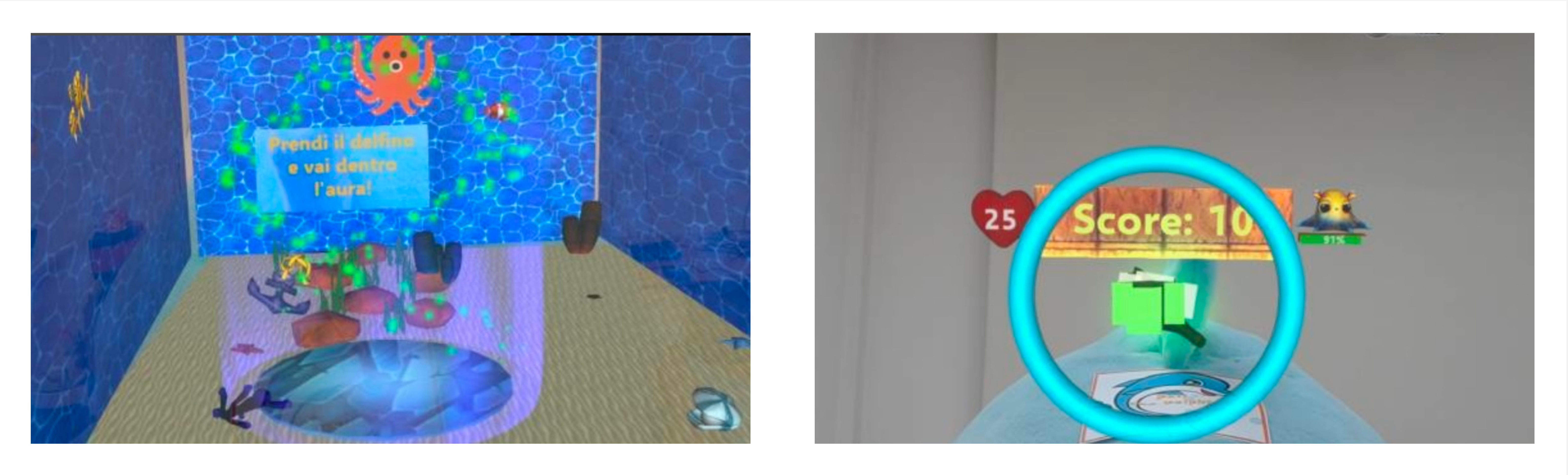}
    \caption{On the left: the room after the spatial mapping transformed in a sea environment. On the right: Infinity Mode; the user correctly collected a ring}
    \label{fig:atofapp}
\end{figure}

ATOF surrounds the end-user with world-anchored virtual elements while maintaining a connection with reality. Users interact with the system through a dolphin-shaped Smart Object, completing tasks such as collecting rings, avoiding obstacles, and opening treasure chests. These tasks take place in a transformed, ocean-themed room filled with interactive digital decorations. The application begins by scanning and transforming the physical room into a virtual ocean floor, adding decorations like sand, water, and sea creatures (Figure \ref{fig:atofapp} left). Users can explore this environment or engage in two game modes: Infinity Mode and Adventure Mode. In Infinity Mode, users interact with a ring generator that produces virtual bubble rings. The user must collect these rings using the Smart Dolphin, ensuring correct angulation as guided by an arrow within each ring (Figure \ref{fig:atofapp} right). Users also avoid obstacles, which decrease health upon collision. Parameters such as obstacle difficulty, duration, and additional tasks can be customized. This mode can run indefinitely until the dolphin’s health is depleted.
Adventure Mode is a story-driven experience where a young dolphin embarks on a journey across the oceans. Users guide the dolphin by completing levels of increasing difficulty, divided into two parts: part one is similar to Infinity Mode, where users can collect bubble rings while avoiding obstacles, while part two is a treasure hunt where users must locate a key emitting spatialized sounds to unlock a chest containing a map to continue the adventure.
The ATOF application was developed using Unity, Visual Studio, Vuforia, and the Mixed Reality Toolkit (MRTK), incorporating spatial mapping to adapt the virtual environment to the room’s dimensions. To enhance performance, virtual elements are preloaded at startup rather than being generated during runtime.

\subsection{Participants} The study involved 11 participants with varying Neurodevelopmental Disorders. The group was balanced in terms of gender, with 5 males (45.5\%) and 6 females (54.5\%). The average age of participants was 30.09 years (SD = 6.24), ranging from 20 to 44 years old. In terms of diagnoses, the majority of participants (45.5\%) had a diagnosis of moderate mental retardation, followed by 27.3\% with severe mental retardation, 18.2\% with genetic syndromes, and 9.1\% diagnosed with an Autism Spectrum Disorder (ASD). The classification of severity was based on DSM-5 guidelines \cite{dsm}. The moderate intellectual disability group typically required support for tasks involving academic and daily living skills, as well as assistance in understanding and interpreting social cues. Participants with severe impairments often needed more substantial support from caregivers during the sessions. All participants and their families were informed about the study’s goals, procedures, and data-handling processes. Written informed consent was obtained from parents or legal guardians. The study adhered to strict ethical guidelines, with approval from the day-care center and compliance with the GDPR standards. 
\subsection{Research Goals and Questionnaires}
The exploratory study aimed to evaluate two key features of the ATOF application: i)
Likeability: how much the participants enjoyed the application, and ii) Usability: the ease of use of the application and its interaction mechanisms.
To assess these features, two questionnaires were designed (Table \ref{tab:likability},\ref{tab:usability}. Questions were answered through a 7-point Likert scale, where 1 means “not much” and 7 means “very much”. Some questions were also presented inverted, as highlighted in Table \ref{tab:likability},\ref{tab:usability}.
\begin{table}[ht]
    \centering
    \begin{tabular}{|c|l|}
        \hline
        \textbf{\#} & \textbf{Likeability Questionnaire} \\ 
        \hline
        Q1 & How much did you like the game? \\ 
        Q2 & Is the game visually appealing? \\ 
        Q3 & Is the dolphin unpleasant? (inverted question) \\ 
        Q4 & Is the game fun? \\ 
        Q5 & Is playing with the bubble rings annoying? (inverted question) \\ 
        Q6 & How much did you enjoy the activity? \\ 
        Q7 & Is the sea environment visually appealing? \\ 
        Q8 & Was the exercise annoying? (inverted question) \\ 
        Q9 & Would you enjoy other games using Mixed Reality? \\ 
        Q10 & Was the game engaging? \\ 
        \hline
    \end{tabular}
    \caption{Likeability Questionnaire}
    \label{tab:likability}
\end{table}

\begin{table}[ht]
    \centering
    \begin{tabular}{|c|l|}
        \hline
        \textbf{\#} & \textbf{Usability Questionnaire} \\ 
        \hline
        Q1 & Was the game easy to play? \\ 
        Q2 & Is the device comfortable to wear? \\ 
        Q3 & Is it easy to hold the dolphin? \\ 
        Q4 & Was it fun to play with both the dolphin and the device? \\ 
        Q5 & Is wearing the device annoying? (inverted question) \\ 
        Q6 & Did you see the projections clearly? \\ 
        Q7 & Were the tasks clear? \\ 
        Q8 & Was it simple to locate the aura? \\ 
        Q9 & Was it easy to find the indicated objects? \\ 
        Q10 & Was the game too tiring? (inverted question) \\ 
        \hline
    \end{tabular}
    \caption{Usability Questionnaire}
    \label{tab:usability}
\end{table}
  
\subsection{Procedure}
Both sessions took place in the gym of the centre hosting the user study. It was a room of approximately 5*4 meters. During the first day, two activities were scheduled in order to make the users familiarise themselves with the application. The experiences were Room Exploration and Standard Tutorial. In the Room Exploration, users were asked to explore the "augmented room" that was completely transformed into an ocean floor. Users could also interact with the virtual decorations (fishes, whales, jellyfish, and more) attached to the walls. They started moving when users approached them, and freeze when the users were far. After the room exploration, each user was asked to play the normal tutorial. In the tutorial, the user had to obtain three rings coming from the center, three rings rotated slightly to the right, and three rings rotated slightly to the left.
The bubble rings kept coming until the user got them all correctly. Following this strategy, users were able to rehearse numerous times how to move correctly to acquire the rings, getting a reward when they succeeded.
The second session was entirely planned to let users try the other tutorial along with the standard one.  In the second tutorial, users had to get one ring coming from the center, one ring coming from the right, and one ring coming from the left. At the end of the second session, participants completed a survey to evaluate the application’s likeability and usability.
\subsection{Results and Discussion}
During the experimentation sessions it was observed interest from most of the users. Some of them wanted to try different task configurations, demonstrating that an application customizable according to the user's preference is a good way to go, especially for users with physical and mental disabilities.
The observations suggest that HoloLens can be a good tool to create new types of experiences for NDD people, where exercises can improve their condition while entertaining them. During the second day of the study, participants had to fill out the questionnaire after having completed the experience.
According to the positive feedback, the trial was both informative and amusing for both users and therapists. The users' enthusiasm in the game was palpable, even when the time was over, some users wanted to play again. Some individuals claimed to have been teleported to their gym from an alternate humorous world.
All of the users accepted the Hololens without any problems, and the fact that they did not lose contact with reality may have helped some of them not to get confused or afraid.
The results regarding the Likeability and Usability parameters are reported in Tables \ref{tab:likeability-atof},\ref{tab:usability-atof}.
\begin{table}[ht]
\centering
\begin{tabular}{|l|l|l|l|l|l|l|l|l|l|l|l|}
\hline
\multicolumn{2}{|l|}{} & \textbf{Q1} & \textbf{Q2} & \textbf{Q3} & \textbf{Q4} & \textbf{Q5} & \textbf{Q6} & \textbf{Q7} & \textbf{Q8} & \textbf{Q9} & \textbf{Q10} \\ \hline
\multicolumn{2}{|l|}{\textbf{Mean Value}} & 7 & 7 & 1 & 7 & 1 & 7 & 7 & 1 & 7 & 7 \\ \hline
\multicolumn{2}{|l|}{\textbf{Minimum}} & 4 & 4 & 1 & 4 & 1 & 6 & 1 & 1 & 1 & 4 \\ \hline
\multicolumn{2}{|l|}{\textbf{Maximum}} & 7 & 7 & 7 & 7 & 6 & 7 & 7 & 7 & 7 & 7 \\ \hline
\end{tabular}
\caption{Likeability results}
\label{tab:likeability-atof}
\end{table}
\begin{table}[ht]
\centering

\begin{tabular}{|l|l|l|l|l|l|l|l|l|l|l|l|}
\hline
\multicolumn{2}{|l|}{\textbf{}} & \textbf{Q1} & \textbf{Q2} & \textbf{Q3} & \textbf{Q4} & \textbf{Q5} & \textbf{Q6} & \textbf{Q7} & \textbf{Q8} & \textbf{Q9} & \textbf{Q10} \\ \hline
\multicolumn{2}{|l|}{\textbf{Mean Value}} & 7 & 7 & 7 & 7 & 1 & 7 & 6 & 4 & 7 & 1 \\ \hline
\multicolumn{2}{|l|}{\textbf{Minimum}} & 2 & 1 & 2 & 4 & 1 & 3 & 4 & 1 & 2 & 1 \\ \hline
\multicolumn{2}{|l|}{\textbf{Maximum}} & 7 & 7 & 7 & 7 & 6 & 7 & 7 & 7 & 7 & 6 \\ \hline
\end{tabular}
\caption{Usability results}
\label{tab:usability-atof}
\end{table}
Only a descriptive analysis of the data was carried out; no additional inferential analysis was possible. However, the results from the descriptive analysis were encouraging, suggesting that the ATOF application was well-received and easy to use. The likeability and usability ratings further support this conclusion, with an average likeability score of M = 5.2 (SD = 1.21) and an average usability score of M = 5.4 (SD = 1.36). These values indicate a generally positive perception of the system, though variability in responses suggests that individual user experiences differed. Because of the limited training provided to this user sample, the outcomes of this exploratory empirical study cannot be generalized; therefore, the findings do not offer conclusive evidence of the benefits that ATOF tasks and the TAR interaction paradigm might bring to individuals with NDD. Additionally, the participants’ diverse ages and diagnostic conditions made thorough data analysis more complex. Nonetheless, experts unanimously acknowledged the potential of this novel paradigm to enhance motor and cognitive abilities.

Because of the limited training provided to this user sample, the outcomes of this exploratory empirical study cannot be generalized; therefore, the findings do not offer conclusive evidence of the benefits that ATOF tasks and the TAR interaction paradigm might bring to individuals with NDD. Additionally, the participants’ diverse ages and diagnostic conditions made thorough data analysis more complex. Nonetheless, experts unanimously acknowledged the potential of this novel paradigm to enhance motor and cognitive abilities. Only a descriptive analysis of the data was carried out; no additional inferential analysis was possible. However, the results from the descriptive analysis were encouraging, suggesting that the ATOF application was well-received and easy to use. Based on the study’s findings, we identified key guidelines for the effective use of the application, taking into account both the generally positive reception reflected in the likeability and usability scores and the variability in individual user experiences.
\begin{itemize}
    \item \textbf{Lesson 1: Target Group.} Participants with severe impairments required more therapist assistance, but some succeeded in completing tasks. This indicates that TAR applications may be most beneficial for users with moderate impairments, though they also show promise for those with more severe challenges.
    \item \textbf{Lesson 2: Customization.} Customization is essential for therapeutic applications in the NDD field. By tailoring tasks to individual abilities, the system can enhance usability and likeability. A user-friendly interface should allow therapists to adjust parameters and track progress over time.
    \item \textbf{Lesson 3: Shared Experience.} Shared experiences, where multiple users and therapists participate, enhance engagement. The ability for therapists to monitor the user’s experience in real-time and the presence of tangible feedback from the Smart Object contributed to the positive group dynamic. Additionally, the physical presence of others may help prevent isolation and maintain awareness of the physical environment.
\end{itemize}

\section{Second Study - Holomarket}
HoloMarket is a Tangible Augmented Reality application designed for persons with NDD. The goal of the application is to teach users how to shop in a supermarket. Holomarket was tested in an exploratory study involving 5 persons with NDD. The aim of the study was to assess the pleasantness and usability of the application (including interaction with the cash register).
\subsection{Application Overview}
\begin{figure}[ht]
    \centering
    \includegraphics[width=\linewidth]{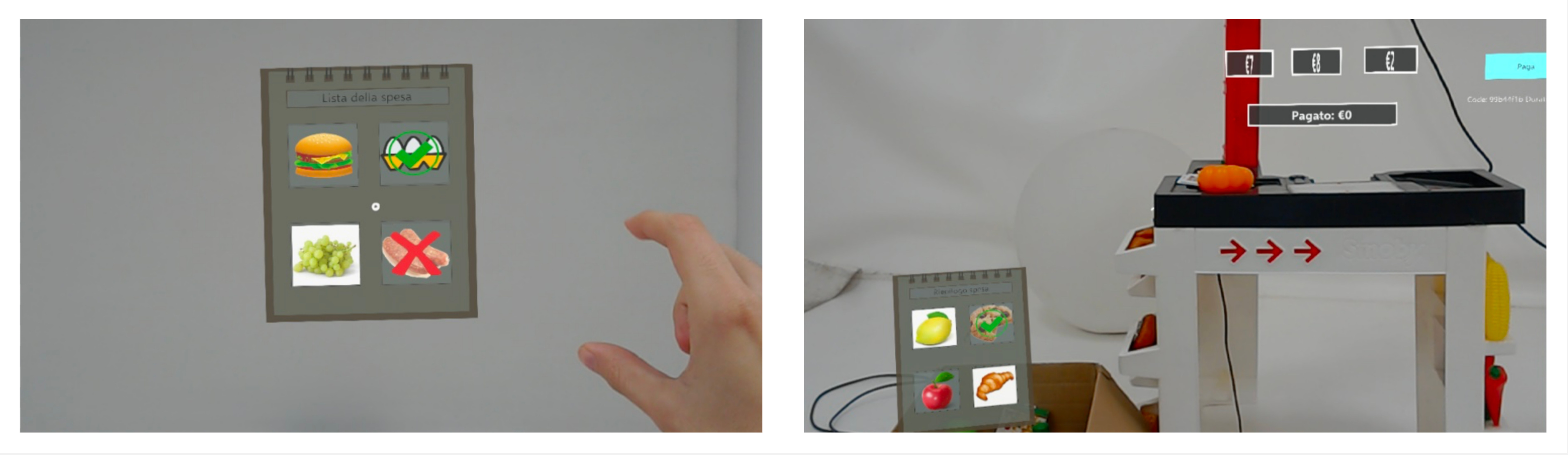}
    \caption{On the left: A green check is shown if food is released in front of a correct image; otherwise, a red cross appears. On the right: correct food passed in Each Price Mode with price text boxes initialized}
    \label{fig:holoapp}
\end{figure}

Holomarket is the product of a co-design process involving caregivers, therapists, psychologists, and engineers. While the detailed co-design process is outside the scope of this paper, the key results relevant to the application are discussed. The requirements for Holomarket were gathered through workshops and interviews with six specialists (one psychologist, three caregivers, and two therapists). This process helped identify tasks best suited to the TAR paradigm and most beneficial for the target population. 
The game design focused on creating a simple, efficient therapy tool. Specialists recommended minimizing distractions and focusing on essential elements like shelves, food, a shopping list, and a cash register. Holographic elements, such as the shopping list and food on the shelves, ensured compatibility with various environments, while physical objects like the cash register and RFID-equipped food items provided tactile feedback. This combination allowed users to practice realistic shopping tasks in a safe, controlled environment.
After starting the application on the Microsoft HoloLens and activating the smart cash register, users choose between two game modes: Each Price Mode and Total Price Mode. Both modes begin with a spatial scan of the room, allowing the system to place virtual shelves and food items appropriately.
Each Price Mode incorporates a memory game into the shopping experience: users search for food items listed on the shopping list. By gazing at an item, they can drag and drop it onto the corresponding image on the list. Correct matches trigger a green check mark; incorrect matches display a red cross (Figure \ref{fig:holoapp} left). After collecting items, users interact with the smart cash register, which shows three price options for each item. Users must select the correct price and pay using RFID-equipped coins or banknotes. If the amount is correct, a green check mark confirms success; otherwise, a red cross indicates an error (Figure \ref{fig:holoapp} right).
In Total Price Mode, the goal is to calculate the total cost of the shopping trip: Users collect items from the shelves, with prices displayed alongside each item and on the shopping list. After gathering items, users pass them through the smart cash register. They must calculate the total cost and pay the correct amount using RFID-enabled banknotes. Errors result in a red cross, while correct payments are confirmed with a green check mark.
\subsection{Participants}
The study involved 5 participants with mild Cognitive Disability, all attending a day-care center for individuals with disabilities with the aim of enhancing personal autonomy and social skills. The group was composed of 2 males (40\%) and 3 females (60\%). The average age of participants was 36.0 years (SD = 6.2). The classification of disability severity was based on DSM-5 guidelines \cite{dsm}. Individuals with mild cognitive disability typically require moderate support in daily living tasks and social interactions. Participants were selected on a volunteer basis. All participants and their families were informed about the study’s goals, procedures, and data-handling processes. Written informed consent was obtained from parents or legal guardians where applicable. The study followed strict ethical guidelines, received approval from the day-care center, and adhered to GDPR standards.
\subsection{Research Goals and Questionnaires}
The primary goal of this study was to explore whether the TAR paradigm, specifically as implemented in Holomarket, was applicable and enjoyable for individuals with NDD. While the theoretical potential of TAR was well-supported by specialists involved in its co-design, this study sought to assess whether the experience was intuitive, clear, and engaging for the intended target population. A positive and pleasant experience was deemed essential, as frustration or confusion could undermine the system’s value as a future training tool. The study investigated two main aspects: likeability (how much the application and its components were enjoyed) and usability (how intuitive and easy the interaction was). The goal was to determine if the combination of virtual and tangible elements in Holomarket could effectively engage users without being overly complex or confusing. Data were collected both qualitatively and quantitatively. Observations were recorded manually during the sessions, documenting participants’ behavior, reactions, and challenges. Quantitative data were gathered using questionnaires completed by participants with the aid of a tablet. These surveys assessed both likeability and usability, providing a mix of objective scores and subjective feedback. Out of the five participants, four datasets were considered valid, as one contained inconsistent and contradictory answers. To evaluate likeability, the study examined how the integration of holograms, physical objects, and gameplay mechanics contributed to the participants’ enjoyment. Specific questions aimed to determine whether the virtual supermarket environment, interactive elements, and overall experience were fun and engaging. Usability was assessed by measuring how intuitive it was for users to interact with the application’s interface and elements, such as selecting items from shelves and using the cash register (Table \ref{tab:likeq},\ref{tab:usabq}). These factors were evaluated using custom-designed questionnaires scored on a 7-point Likert scale.
\begin{table}[ht]
\centering
\begin{tabular}{|l|l|}
\hline
\textbf{\#} & \textbf{Likeability Questionnaire} \\ \hline
\textbf{Q1} & I really enjoyed the game. \\ \hline
\textbf{Q2} & The game is beautiful. \\ \hline
\textbf{Q3} & The game is fun. \\ \hline
\textbf{Q4} & Playing with foods on the shelves was annoying. \\ \hline
\textbf{Q5} & Playing the activity was fun. \\ \hline
\textbf{Q6} & The supermarket environment was beautiful \\ \hline
\textbf{Q7} & Playing with the cash register and the other objects was funny \\ \hline
\textbf{Q8} & The exercise was boring \\ \hline
\textbf{Q9} & I would like to play other games with Mixed Reality devices. \\ \hline
\textbf{Q10} & The game was engaging? \\ \hline
\end{tabular}
\caption{Likeability questions}
\label{tab:likeq}
\end{table}
\begin{table}[ht]
\centering
\begin{tabular}{|l|l|}
\hline
\textbf{\#} & \textbf{Usability Questionnaire} \\ \hline
\textbf{Q1} & I really liked the game. \\ \hline
\textbf{Q2} & The headset is comfortable to wear. \\ \hline
\textbf{Q3} & Taking objects from the shelves is easy. \\ \hline
\textbf{Q4} & Using the cash register and its objects is easy. \\ \hline
\textbf{Q5} & Using the Mixed Reality device with the cash register is funny? \\ \hline
\textbf{Q6} & Keeping the Mixed Reality device on your head is annoying  \\ \hline
\textbf{Q7} & I can see well what the device projected \\ \hline
\textbf{Q8} & It was always what to do \\ \hline
\textbf{Q9} & It was clear to understand where the foods were on the shelves \\ \hline
\textbf{Q10} & The cash register was easy to use \\ \hline
\textbf{Q11} & The game was too hard? \\ \hline
\end{tabular}
\caption{Usability questions}
\label{tab:usabq}
\end{table}

\subsection{Procedure} The study was performed in one of the centers hosting the participants. In order to recreate an environment as similar as possible to the aisle of a supermarket, it was decided to use the main aisle inside the center. The space occupied by the AR environment was 3x5 m. The cash register was located in front of the user at a distance of 5m. The study consisted of a single session in which all participants, one at a time, tried the \textit{Each Price Mode} scenario. Each Price Mode task was configured with fifteen foods on the shelves and four foods on the shopping list. At the end of each session, the participant answered questions from the likeability and usability questionnaires (Tables \ref{tab:likeq}, \ref{tab:usabq}). 

\subsection{Results and Discussion}
All five people filled out the questionnaires at the end of the test, but one was discarded due to inconsistent and contradictory answers. Tables \ref{tab:like-r}, \ref{tab:use-r} report the answers collected from the participants.

\begin{table}[ht]
\centering
\begin{tabular}{|ll|l|l|l|l|l|l|l|l|l|l|}
\hline
\multicolumn{1}{|c|}{\textbf{Gender}} & \multicolumn{1}{c|}{\textbf{Age}} & \multicolumn{1}{c|}{\textbf{Q1}} & \multicolumn{1}{c|}{\textbf{Q2}} & \multicolumn{1}{c|}{\textbf{Q3}} & \multicolumn{1}{c|}{\textbf{Q4}} & \multicolumn{1}{c|}{\textbf{Q5}} & \multicolumn{1}{c|}{\textbf{Q6}} & \multicolumn{1}{c|}{\textbf{Q7}} & \multicolumn{1}{c|}{\textbf{Q8}} & \multicolumn{1}{c|}{\textbf{Q9}} & \multicolumn{1}{c|}{\textbf{Q10}} \\ \hline
\multicolumn{1}{|l|}{Female} & 32 & 6 & 7 & 7 & 1 & 7 & 7 & 7 & 1 & 7 & 7 \\ \hline
\multicolumn{1}{|l|}{Female} & 32 & 7 & 7 & 7 & 1 & 7 & 7 & 7 & 1 & 7 & 6 \\ \hline
\multicolumn{1}{|l|}{Male} & 47 & 4 & 7 & 7 & 5 & 7 & 7 & 5 & 1 & 7 & 7 \\ \hline
\multicolumn{1}{|l|}{Female} & 34 & 5 & 6 & 7 & 5 & 7 & 7 & 5 & 4 & 7 & 7 \\ \hline
\multicolumn{2}{|l|}{\textbf{Mean Value}} & 5,5 & 6,75 & 7 & 3 & 7 & 7 & 6 & 1,75 & 0 & 6,75 \\ \hline
\multicolumn{2}{|l|}{\textbf{Standard Deviation}} & 1,29 & 0,5 & 0 & 2,31 & 0 & 0 & 1,15 & 1,5 & 0 & 0,5 \\ \hline
\end{tabular}
\caption{Likeability results}
\label{tab:like-r}
\end{table}

\begin{table}[ht]
\centering
\begin{tabular}{|ll|l|l|l|l|l|l|l|l|l|l|l|}
\hline
\multicolumn{1}{|c|}{\textbf{Gender}} & \multicolumn{1}{c|}{\textbf{Age}} & \multicolumn{1}{c|}{\textbf{Q1}} & \multicolumn{1}{c|}{\textbf{Q2}} & \multicolumn{1}{c|}{\textbf{Q3}} & \multicolumn{1}{c|}{\textbf{Q4}} & \multicolumn{1}{c|}{\textbf{Q5}} & \multicolumn{1}{c|}{\textbf{Q6}} & \multicolumn{1}{c|}{\textbf{Q7}} & \multicolumn{1}{c|}{\textbf{Q8}} & \multicolumn{1}{c|}{\textbf{Q9}} & \multicolumn{1}{c|}{\textbf{Q10}} & \multicolumn{1}{c|}{\textbf{Q11}} \\ \hline
\multicolumn{1}{|l|}{Female} & 32 & 5 & 7 & 1 & 7 & 7 & 7 & 7 & 7 & 3 & 7 & 1 \\ \hline
\multicolumn{1}{|l|}{Female} & 32 & 5 & 6 & 4 & 5 & 7 & 2 & 7 & 7 & 7 & 5 & 1 \\ \hline
\multicolumn{1}{|l|}{Male} & 47 & 4 & 4 & 4 & 5 & 7 & 4 & 7 & 7 & 7 & 5 & 1 \\ \hline
\multicolumn{1}{|l|}{Female} & 34 & 2 & 7 & 3 & 3 & 4 & 6 & 4 & 4 & 4 & 3 & 6 \\ \hline
\multicolumn{2}{|l|}{\textbf{Mean Value}} & 4 & 6 & 3 & 5 & 6,25 & 4,75 & 6,25 & 6,25 & 5,25 & 5 & 2,25 \\ \hline
\multicolumn{2}{|l|}{\textbf{Standard Deviation}} & 1,41 & 1,41 & 1,41 & 1,63 & 1,5 & 2,22 & 1,5 & 1,5 & 2,06 & 1,63 & 2,5 \\ \hline
\end{tabular}
\caption{Usability results}
\label{tab:use-r}
\end{table}
The study results suggest that overall interactions with Holomarket were positive, particularly in terms of engagement and enjoyment. The average likeability score was 51.75 out of 70 (SD = 6.94), indicating that most participants found the experience entertaining and immersive. Notably, participants appreciated the virtual supermarket environment and the link between the shopping list and items on the shelves. This combination of real and virtual elements created a pleasant environment, where users felt comfortable and did not experience disorientation upon wearing the HoloLens. However, several challenges emerged, especially in object interaction. Many participants struggled with the pinching gesture needed to select and move virtual food items, suggesting that this action might not be intuitive for all users; external assistance was often required to progress. In addition, the system feedback was occasionally unclear, making it difficult for participants to determine whether they had completed a task correctly or how to move forward. Usability scores were also high, with an average of 63.5 out of 70 (SD = 5.68), reflecting that participants generally found the system intuitive and easy to navigate. Nevertheless, some usability issues arose, notably regarding the HoloLens’ wearability. Many participants frequently adjusted the device or needed breaks due to discomfort. The limited field of view further affected interaction, at times making it hard to see all necessary elements in the virtual space. These findings underscore the promise of TAR applications like Holomarket as engaging therapeutic tools for individuals with Neurodevelopmental Disorders, fostering skill development in an immersive and enjoyable manner. At the same time, the study highlights key usability challenges, particularly in terms of interaction mechanics, system feedback, and device wearability. For instance, the difficulties with pinching gestures suggest a need for alternative interaction methods (e.g., larger activation zones or simpler selection mechanics). Similarly, the lack of clear system feedback—especially in cash register interactions—indicates the importance of improved visual, auditory, or haptic cues. Lastly, discomfort with the HoloLens, partly due to its limited field of view, points to the need for exploring alternative AR hardware or software adjustments to accommodate diverse user needs. Despite these challenges, Holomarket appears to have strong potential for enhancing autonomy, cognitive skills, and overall engagement in individuals with NDD. 

\section{Conclusion}
This work presents an innovative interactive paradigm designed to address challenges faced by individuals with Neurodevelopmental Disorders in their everyday lives. The solution leverages Tangible Augmented Reality, a paradigm that integrates Augmented Reality devices with Smart Objects/Tangible Interface), providing tangible feedback to create realistic simulations of everyday scenarios. To our knowledge, these are the first studies exploring the application of TAR for individuals with NDD. Two applications have been developed under this paradigm: Holomarket and Along the Oceanic Flow (ATOF). Holomarket replicates a supermarket environment, enabling users to learn grocery shopping tasks such as selecting items from shelves, using a shopping list, and completing payments with a real smart cash register. ATOF transforms a user’s room into an oceanic environment where users interact with a dolphin-shaped Smart Object to complete tasks that improve motor and cognitive skills.
The development of TAR applications, particularly for NDD populations, was challenging. A lack of established design principles for TAR applications required iterative development, with feedback from therapists and specialists informing multiple design revisions. Establishing seamless communication between the AR device (Microsoft HoloLens) and Smart Objects, such as the cash register in Holomarket, posed technical hurdles. Additionally, usability challenges arose from the limited field of view of the HoloLens and the unnatural gesture controls, which participants often struggled to master. Prolonged use of the HoloLens also caused discomfort for some participants, with the device’s pads proving inadequate to ensure comfort during extended sessions.
Despite these challenges, the exploratory studies demonstrated promising results. Participants enjoyed using TAR applications, finding the simulations engaging and motivating. Both Holomarket and ATOF successfully captured the interest of users and encouraged them to continue practicing tasks.
Participants faced difficulties with certain interactions, such as grasping and moving virtual objects, as well as interpreting interface feedback during tasks like payment in Holomarket. Nevertheless, therapists and participants highlighted the potential of TAR to enhance skills while providing a safe and enjoyable experience.

\subsection{Limitations and Future Work}
We acknowledge some limitations in our studies. The small sample size limited the ability to apply statistical methods or generalize findings. The short duration of the studies restricted the assessment of long-term benefits or improvements. Microsoft HoloLens also presented constraints, such as its limited field of view, which restricted spatial awareness, and its discomfort during prolonged use, which affected participant engagement. Additionally, many users found the gesture-based interactions challenging, particularly the pinch-and-drag motion, which felt unnatural and difficult to master. Future work will focus on enhancing the technology and conducting more rigorous studies. Improving data collection capabilities within TAR applications will allow for detailed performance analysis, including metrics such as task completion time and error rates. Adding features to enable caregivers to view the user’s perspective in real time will further enhance the usability and effectiveness of these applications.
Upgrading to Microsoft HoloLens 2 will simplify interactions through improved gesture recognition, enabling more intuitive control of virtual objects. Collaborative experiences could also be explored, such as coupling two HoloLens devices with different Smart Objects to enable cooperative problem-solving tasks.
Longer-term research will focus on conducting a six-month controlled study to evaluate the effectiveness of TAR applications for improving motor and cognitive skills in individuals with NDD. Collaborations with three therapeutic centers in Milan will provide a larger sample size to address these goals. These efforts aim to compare TAR-based interventions with traditional therapies and establish the paradigm’s efficacy.

\begin{credits}

\subsubsection{\ackname} This research was carried out within the MUSA – Multilayered Urban Sustainability Action – project, funded by the European Union – NextGenerationEU, under the National Recovery and Resilience Plan (NRRP) Mission 4 Component 2 Investment Line 1.5: Strengthening of research structures and creation of R\&D “innovation ecosystems”, set up of “territorial leaders in R\&D".

In this paper, we used Overleaf’s built-in spell checker, the current version of ChatGPT (GPT 4.0), and Grammarly. These tools helped us fix spelling mistakes and get suggestions to improve our writing. If not noted otherwise in a specific section, these tools were not used in other forms.
\end{credits}
%
%
%
%

\end{document}